\documentclass[12pt]{iopart}

\usepackage{iopams}  

\usepackage{times}
\usepackage{bm}
\usepackage{bbm}
\usepackage{graphicx}
\usepackage{epsfig}
\usepackage{latexsym}
\usepackage{nicefrac}
\def\corresponds{{\lower.2ex\hbox{=}}{\rm\kern-.75em^\triangle}}
\def\succsim{\succ\kern-.9em_\sim\kern.3em}
\def\precsim{\prec\kern-1em_\sim\kern.3em}
\def\slantfrac#1#2{\kern1em^{#1}\kern-.3em/\kern-.1em_{#2}}
\def\lfrac#1#2{{}^{#1\!}\kern-.0em/_{#2}}

\def\buildrel#1\under#2{\mathrel{\mathop{\kern0pt #2}\limits_{#1}}}

\newcommand{\hf} {\frac{1}{2}}

\newcommand{\uvarphi}{{\underline \varphi}}

\def\ord#1{{\mathcal O}(#1)}

\def\mr#1{{\mathrm{#1}}}

\def\cL{{\mathcal L}}

\begin{document}

\title[Symmetries and Phase Structure of the Layered Sine-Gordon Model]
{Symmetries and Phase Structure of the Layered Sine-Gordon Model}

\author{I. N\'andori}

\address{Institute of Nuclear Research of the Hungarian Academy of
Sciences, Debrecen, H-4001 Debrecen, P.O.Box 51, Hungary}
\ead{nandori@atomki.hu}

\begin{abstract}
The phase structure of the layered sine-Gordon (LSG) model is 
investigated in terms of symmetry considerations by means of  
a differential renormalization group (RG) method, within the local 
potential approximation. The RG analysis of the general $N$-layer 
model provides us with the possibility to consider the dependence 
of the vortex dynamics on the number of layers. The Lagrangians 
are distinguished according to the number of zero eigenvalues of 
their mass matrices. The number of layers is found to be decisive with 
respect to the phase structure of the $N$-layer models, with 
neighbouring layers being coupled by terms quadratic in the field 
variables. It is shown that the LSG model with $N$ layers undergoes 
a Kosterlitz--Thouless type phase transition at the critical value 
of the parameter $\beta^2_{\rm c} = 8 N \pi$. In the limit of 
infinitely many layers the LSG model can be considered as the 
discretized version of the three-dimensional sine-Gordon model 
which has been shown to have a single phase within the local 
potential approximation. The infinite critical value of the 
parameter $\beta_c^2$ for the LSG model in the continuum limit 
($N\to \infty$) is consistent with the latter observation.
\end{abstract}

%Uncomment for PACS numbers title message
% Keywords required only for MST, PB, PMB, PM, JOA, JOB? 
\vspace{2pc}
\noindent{\it Keywords}: Renormalization group method;
Renormalization; Two-dimensional field theory
% Uncomment for Submitted to journal title message
\submitto{\JPA}
% Comment out if separate title page not required
\maketitle

%-----------------------------------------------------------------------
% Introduction
%-----------------------------------------------------------------------
\section{Introduction}
\label{intro} 

The renormalization of sine-Gordon (SG) type models represents a challenge 
in quantum field theory, where the usual strategies are based on the Taylor 
expansion of the interaction Lagrangian. However, in the case of an SG type 
scalar field theory where the self-interaction is given by a periodic term 
(periodic in the field variable), any truncation of the Taylor expansion
of the potential violates the essential symmetry of the model. As it is well 
known, the phase structure of the system crucially depends on the symmetries 
of the interaction Lagrangian in the field variable. Therefore, in order to
obtain the low-energy effective theory and to map out the phase structure 
of a SG type model one has to use a method which retains the periodicity of
the system.  

The phase structure of the "pure" SG model which is periodic in the internal 
space spanned by the field variable has been investigated in great 
detail \cite{Ko1974,Amit,sg_model} and as it is well known, 
the model has two phases separated by the critical value of the parameter, 
$\beta_c^2 = 8\pi$ \cite{coleman,sg2}. Another interesting subject 
concerns the ``massive'' SG model \cite{sanyi,Fi1979}
where the periodicity is broken by the explicit mass term of the Lagrangian. 
The massive SG model has a single phase, all the coupling constants of the model 
are relevant parameters independently of $\beta$. It is an interesting subject 
to consider a SG type theory which combines the ``features'' of the massless
and massive SG models where the periodicity is broken only partially. The 
following generalization of the SG model which is called the layered 
sine-Gordon (LSG) model~\cite{Fi1979,HeHoIs1995} belongs to the latter 
category, 
\begin{equation}
\label{clhere}
\cL_{\rm LSG} = 
{\frac12} \, \sum_{i=1}^{N} (\partial {\varphi_i})^2
+ {\frac12} \,J\, \sum_{i=1}^{N-1} (\varphi_{i+1} - \varphi_{i})^2 +
U(\varphi_1,...,\varphi_N)\,,
\end{equation}
where each $\varphi_i$ is a one-component Lorentz-scalar field and 
the second term corresponds to the coupling between the SG models 
(i.e. layers). $U(\varphi_1,...,\varphi_N)$ is assumed to be 
periodic but the periodicity is broken (partially) by the interlayer 
coupling terms. The LSG model has relevance in high-energy and 
low-temperature physics. The sine-Gordon model with $N$-layers can be
considered as the bosonized version of the $N$-flavour Schwinger 
model~\cite{Fi1979}. Another suitable generalization of the SG 
model is the $SU(N)$ Thirring model~\cite{banks,Halpern}. 
The LSG model with $N=2$ layers has been used to describe the
vortex dominated properties of high-$T_{\rm  c}$ superconductors 
which have a layered structure \cite{pierson_lsg,kalman}.

Recently, the LSG model with two coupled layers, have been analyzed 
in the framework of the non-perturbative Wegner--Houghton (WH) 
renormalization group (RG) method which retains the periodicity 
of the model~\cite{NaEtAl2005}. The WH--RG approach \cite{wh} is 
incompatible with the derivative expansion due to the sharp momentum 
cutoff used. However, it represents one of the most straightforward 
implementations of a functional RG method. As such, it is a rather 
powerful tool for the analysis of the RG flow of theories with 
periodic self-interactions, including situations with more than one 
interacting field and the higher harmonics which may be generated during the 
RG evolution for periodic self-interactions. The WH--RG method provides us 
with a suitable tool for the investigation of the phase structure of these 
periodic field theories. In this paper, we would like to present a further 
contribution to the study of related models by means of the WH--RG 
method performed for the LSG model with $N$ layers. 

We present an explicit rotation in the internal space of the field 
variables, which allows us to decompose the Lagrangians into ``periodic'' 
and ``non-periodic'' fields. In this article we refer to a field 
variable whose self-interaction is characterized by a periodic 
function with and without an explicit symmetry breaking mass term as 
a ``non-periodic'' and ``periodic'' mode, respectively. 

The purpose of this rotation is twofold. First, we would like to compare the
result of our RG analysis to that of the perturbative treatment discussed
in Ref. \cite{plb} where the rotated $N$-layer SG model is studied. In the 
infrared (IR) region, with $k \ll M$ (with $M$ being the mass eigenvalue and 
$k$ is the momentum cutoff), it is allowed to use perturbation theory but 
only for the non-periodic modes \cite{ZJ1996}. Second, by using the 
rotation we would like to demonstrate that the internal symmetry in the 
field variable is decisive for the phase structure of the LSG model. 

The non-periodic modes have a trivial tree level scaling, which is consistent 
with the explicit breaking of the internal periodicity in the field variable. 
In the limit of a vanishing inter-layer coupling $J$, the coupled $N$-layer 
model approaches the sum of $N$ decoupled sine-Gordon models, each of which 
has a Kosterlitz--Thouless type phase transition at the critical value 
$\beta_c^2 = 8 \pi$ (see Refs.~\cite{Ko1974,Amit,sg_model,sg2}). For an $N$-layer 
model with a non-vanishing coupling $J$, we show here that there exists 
exactly one periodic mode which has two types of scaling behaviour
separated by the critical value $\beta_c^2 = 8 N \pi$ where $N$ is the number
of layers.

Our paper is organized as follows. In Sec.~\ref{def}, we define the layered 
SG model and discuss the connection to the two-dimensional (2D) and 
three-dimensional (3D) SG models. We then give basic relations used for 
the Wegner--Houghton RG method~\cite{wh} and derive the mass-corrected 
ultra-violet (UV) WH--RG equation in Sec.~\ref{wh}. In Sec.~\ref{N=2}, 
the flavour-doublet and in Sec.~\ref{N=3}, the flavour-triplet LSG models 
are analyzed by means of the UV mass-corrected WH--RG method in detail.
The  generalization to $N$ layers is also studied. 
In Sec.~\ref{3dsg}, the UV-RG evolution of the 3D-SG model is investigated
and compared to that of the LSG model with $N$ layers.  
Finally, we conclude with a summary in Sec.~\ref{sum}.

%-----------------------------------------------------------------
% Layered sine-Gordon model
%------------------------------------------------------------------
\section{Layered sine-Gordon model}
\label{def} 

The LSG model belongs to a wider class of massive SG type theories
due to the interlayer coupling which can be considered as a mass term. 
In general, the bare Lagrangian for the massive SG model with 
$N$-layers is~\cite{NaEtAl2005,plb}
\begin{equation}
\label{clb}
\cL = 
{\frac12} \sum_{i=1}^N (\partial {\varphi_i})^2
+ {\frac12} \sum_{i,j}^N \varphi_i M_{i,j}^2 \varphi_j
+ U(\varphi_1,...,\varphi_N),
\end{equation}
where $\varphi_i$ is a one-component scalar field, the theory is 
constructed in $d=2$ dimensions in Euclidean metric and the periodic 
self-interaction is given by the term 
\begin{equation}
\label{property2}
U(\varphi_1,...,\varphi_N)= 
U\left(\varphi_1+ \frac{2\pi}{\beta_1},...,
\varphi_N+\frac{2\pi}{\beta_N}\right)\,.
\end{equation}
The model has a global $Z(2)$ discrete symmetry 
$\varphi_i \to -\varphi_i$. By applying an orthogonal transformation 
on the flavour multiplet $(\varphi_1,\varphi_2,\dots,\varphi_N)$, the 
massive SG model transforms into a similar one with transformed period 
lengths in the internal space. Since the global $O(N)$ rotation does 
not mix the field fluctuations with different momenta, the scaling laws 
and the phase structure should be the same for all the rotated models.

The mass matrix $M^2_{ij}$ $(i,j=1,2, ... ,N)$ is symmetric and 
positive semidefinite and assumed to have a special ``interlayer'' 
structure, 
\begin{equation}
\label{massiveLSG}
\cL = 
{\frac12} \sum_{i=1}^{N} (\partial {\varphi}_i)^2
+ {\frac12} \sum_{i=1}^{N} M^2_i \varphi_{i}^2 
+ {\frac12} J \sum_{i=1}^{N-1} (\varphi_{i+1} - \varphi_{i})^2 +
U(\varphi_1,...,\varphi_N),
\end{equation}
where the explicit mass terms are $M^2_i$ ($i=1,...,N$) and $J$ 
describes the interaction between the layers. Since the layers are 
assumed to be equivalent $M_i^2 \equiv M^2$ for ($i=1,...,N$) is a 
natural choice. The symmetries and phase structure of the massive SG 
model (\ref{massiveLSG}) with $N=2$ layers has already been discussed 
in Ref.~\cite{NaEtAl2005}. It was demonstrated that the number of 
zero eigenvalues of the mass matrix is found to be decisive with 
respect to the phase structure of the model. The mass eigenvalues of 
the layered system (\ref{massiveLSG}) for $N=2,3,4$ layers are 
[$M^2, 2J+M^2$], [$M^2, J+M^2, 3J+M^2$] and 
[$M^2, 2J+M^2, (2+\sqrt{2})J+M^2,(2-\sqrt{2})J+M^2$], respectively.
Consequently, for vanishing explicit masses ($M^2=0$) the layered 
model has always a single zero mass eigenvalue and as it was shown 
for the $N=2$-layer case, it undergoes a phase transition. This can 
also be understood in terms of symmetry considerations. In the presence 
(absence) of explicit mass terms the periodic symmetry of the layered 
model (\ref{massiveLSG}) is broken entirely (partially). In this article
we would like to clarify this general statement by considering the
phase structure of the $N$-layer SG model by means of the differential 
RG approach. 

For vanishing explicit mass terms ($M^2=0$) the Lagrangian 
(\ref{massiveLSG}) takes the form of Eq.~(\ref{clhere}),
\begin{equation}
\cL_{\rm NLSG} = 
{\frac12} \, \sum_{i=1}^{N} (\partial {\varphi}_i )^2
+ {\frac12} \,J\, \sum_{i=1}^{N-1} (\varphi_{i+1} - \varphi_{i})^2 +
U(\varphi_1,...,\varphi_N)\,,
\end{equation}
with $\beta_i = \beta$ (for $i=1,2,...,N$). The LSG model with $N=2$ 
layers has been proposed as an adequate description of the vortex 
dominated properties of strongly anisotropic high transition temperature 
superconductors which have a layered structures. In this case the periodic 
term has a simple structure 
\begin{equation}
\label{lsg}
\cL_{\rm 2LSG} = 
{\frac12} \, \sum_{i=1}^{2} (\partial {\varphi}_i )^2
+ {\frac12} \,J\, (\varphi_2 - \varphi_{1})^2 +
u \, [\cos(\beta\varphi_1) + \cos(\beta\varphi_2)] \,,
\end{equation}
where $u$ corresponds to the fugacity parameter of the vortex system, 
$\beta$ is related to the temperature and the second term describes the 
weak Josephson coupling between the superconducting layers 
\cite{pierson_lsg}.

Finally, we would like to demonstrate that in the limit $N\to\infty$ 
the LSG model can be considered as the discretized version of the 3D-SG 
model. The 3D-SG model has the following action  
\begin{equation}
S = \int d^3 r \left[ {\frac12} (\partial_{\mu} \varphi_{3D})^2 
+ u_{3D} \cos(\beta_{3D}\varphi_{3D})\right],
\end{equation}
where $\varphi_{3D}\equiv\varphi_{3D}(x,y,z)$ is a one-component scalar 
field and $\beta_{3D}$, $u_{3D}$ are the dimensionful parameters of the 
theory. The model is constructed in $d=3$ spatial dimensions with an 
Euclidean metric. The anisotropic 3D-SG model reads as
\begin{equation}
\label{aniso-3dsg}
S = \int d^3 r \left[ \frac{1}{2\beta^2_{\parallel}}
[(\partial_{x} \theta)^2 + (\partial_{y} \theta)^2]  
+ \frac{1}{2\beta^2_{\perp}} (\partial_{z} \theta)^2
+ u_{3D} \cos(\theta)\right],
\end{equation}
where $\theta = \varphi_{3D} \beta_{3D}$ is introduced. In the isotropic
limit $\beta_{\parallel} = \beta_{\perp} \equiv \beta_{3D}$ is assumed.
Rescaling the field $\Phi = \theta / \beta_{\parallel}$, the action 
(\ref{aniso-3dsg}) becomes
\begin{equation}
\label{modif-3dsg}
S = \int d^3 r \left[ \frac{1}{2}
[(\partial_{x} \Phi)^2 + (\partial_{y} \Phi)^2]  
+ \frac{\beta^2_{\parallel}}{2\beta^2_{\perp}} (\partial_{z} \Phi)^2
+ u_{3D} \cos(\beta_{\parallel}\Phi)\right].
\end{equation}
In case of very strong anisotropy, the continuous derivation and the 
integration in the $z$-direction is replaced by finite difference and 
summation, respectively, 
\begin{equation}
\partial_z {\Phi}(x,y,z)\to 
\frac{\Phi(x,y,z+s)-\Phi(x,y,z)}{s}, 
\hskip 1cm
\int dz \to \sum_{z=1}^N s,
\end{equation}
where $s$ is the interlayer distance. Using this discretization, one 
arrives at the LSG model with $N$ layers
\begin{equation}
S =  \int d^2 r \left[{\frac12} \sum_{i=1}^N (\partial \varphi_i)^2   
+ \frac{1}{2} \, J \, \sum_{i=1}^{N-1} (\varphi_{i+1} - \varphi_{i})^2 
+ u  \sum_{i=1}^N \cos(\beta\varphi_i)\right] \,,
\end{equation}
where $\varphi_i(x,y)\equiv \sqrt{s} \Phi(x,y,z=i)$,
$J \equiv \beta^2_{\parallel}/ (\beta^2_{\perp} s^2) $, 
$\beta\equiv\beta_{\parallel}/\sqrt{s}$ and $u\equiv s u_{3D}$ are 
introduced. Therefore, in the continuum limit $N\to \infty$ the LSG 
model can be considered as the discretized version of the 3D-SG model 
and for $N=1$ the LSG model reduces to the 2D-SG model.

%-------------------------------------------------------------------
% Wegner--Houghton RG method
%--------------------------------------------------------------------
\section{Wegner--Houghton renormalization group method}
\label{wh}

In order to map out the phase structure of the layered system, we 
perform an RG analysis for the layered SG model by means of the 
differential RG approach in momentum space where the blocking 
transformations are realized by successive elimination of the field 
fluctuations according to their decreasing momentum in infinitesimal 
steps \cite{Wi1971}. The high-frequency modes are integrated out above 
the moving momentum cutoff $k$ and the physical effects of the 
eliminated modes are encoded in the scale-dependence of the coupling 
constants. The elimination of the modes above the moving scale $k$ is 
complete in Wegner's and Houghton's method (WH--RG) \cite{wh} because 
of the sharp momentum cutoff. The WH method provides a functional RG 
equation for the blocked action. In order to solve the WH--RG equation, 
one has to project it to a particular functional subspace. Therefore, 
one generally assumes that the blocked action contains only local 
interactions, then let expand it in powers of the gradient of the field 
and truncate this expansion at a given order, for technical reasons 
\cite{janosRG}. Here we restrict ourselves to the leading order of 
the gradient expansion, i.e. to the local-potential approximation 
(LPA). The blocked action for the LSG model with $N$-layers reads as
\begin{equation}
S_k = \int d^2x \left[\frac{1}{2} \sum_{i=1}^N (\partial \varphi_i)^2
+ V_k(\varphi_1,...,\varphi_N)\right],
\end{equation}
where $k$ is the running momentum cutoff and 
$V_k(\varphi_1,...,\varphi_N)$ is the blocked potential which has the
following form  
\begin{equation}
\label{ansatz1}
V_k(\varphi_1,...,\varphi_N) = 
{\frac12}{ \uvarphi}^\mr{T} {\underline {\underline M}}^2_k { \uvarphi}
+ U_k(\varphi_1,...,\varphi_N),
\end{equation}
where $U_k$ is the periodic part of the blocked potential and 
${\underline{\underline M}}^2_k$ represents the scale-dependent mass 
matrix. Notice, that the momentum scale-dependence is encoded in the 
coupling constants of the model. The WH--RG equation in LPA has been 
derived for two interacting scalar fields in 
Refs.~\cite{NaEtAl2005,philmag}. The generalization for the $N$-layer 
SG model is straightforward and can be written as
\begin{equation}
\label{dimful_wh}
k \, \partial_k V_k(\varphi_1,...,\varphi_N) = 
- \frac{k^2}{4\pi} 
\ln \left(\frac{\det[\delta_{ij} k^2 + V^{ij}_k]}{k^{2N}} \right)\,,
\end{equation}
where $\delta_{ij}$ is the Kronecker delta, $V^{ij}_k = 
\partial_{\varphi_i}\partial_{\varphi_j}V_k (\varphi_1,...,\varphi_N)$ 
is the second derivative of the dimensionful blocked potential with 
respect to the field variables. We then introduce dimensionless parameters
in order to remove the trivial scale-dependence of the coupling constants. 
The WH--RG equation in LPA for the dimensionless blocked potential reads as
\begin{equation}
\label{dimless_wh}
(2 + k \, \partial_k) {\tilde V}_k(\varphi_1,...,\varphi_N) = 
- \frac{1}{4\pi} 
\ln \left(\det[\delta_{ij} + {\tilde V}^{ij}_k] \right)\,,
\end{equation}
where $\tilde V_k = k^{-2} V_k$ is introduced. All dimensionless 
quantities will be denoted with a tilde superscript in the following.
We recall that in $d=2$ dimensions the scalar fields carry no 
physical dimension, so that $\varphi_i=\tilde\varphi_i$ and hence 
$\beta = \tilde\beta$.

Inserting the dimensionless form of the ansatz (\ref{ansatz1}) into the 
WH--RG equation (\ref{dimless_wh}), the right hand side turns out 
to be periodic, while the left hand side contains both periodic and
non-periodic parts. The non-periodic part contains the mass term 
and we obtain the trivial tree-level evolution for the dimensionless 
mass parameters ${\tilde M}^2_{ij}(k)$,
\begin{equation}
\label{treelevel}
{\tilde M}^2_{ij}(k) =
{\tilde M}^2_{ij} (\Lambda) \left({k\over\Lambda}\right)^{-2}
\end{equation}
where ${\tilde M}^2_{ij} (\Lambda)$ is the initial value for the mass 
term at the UV momentum cutoff $\Lambda$. Therefore, the dimensionful
mass terms have no evolution, i.e. $M^2_{ij}$ are scale-independent. 
The RG flow equation
\begin{equation}
\label{Uflow}
(2 + k \, \partial_k) {\tilde U}_k (\varphi_1,...,\varphi_N) 
= - \frac{1}{4 \pi} 
\ln\left(\det[\delta_{ij} + {\tilde V}^{ij}_k]\right) \,
\end{equation}
stands for the dimensionless periodic piece of the blocked potential.

In order to obtain the scale-dependence of the coupling constants, 
one has to solve the differential equation (\ref{Uflow}) which
can be done only numerically. However, analytic solutions are available
by considering asymptotic approximations of Eq.~(\ref{Uflow}). In
case of the UV approximation, the potential is assumed to be much smaller
then the momentum cutoff $k$ and the logarithm can be linearized
in Eq.~(\ref{Uflow}). In order to obtain reliable UV scaling laws 
which can be used to determine the phase structure of the layered model, 
one has to incorporate the effect of the mass terms (coupling between 
the layers). Therefore, one should use the ``mass-corrected'' UV 
approximation of Eq.~(\ref{Uflow}) which has been discussed in 
\cite{NaEtAl2005}. We here briefly summarize the derivation of the 
mass-corrected UV-RG where the argument of the logarithm is expanded
in powers of $\tilde U_k$    
\begin{equation}
\det[\delta_{ij} + {\tilde V}^{ij}_k] \approx
C + F_1(\tilde U_k) + F_2(\tilde U^2_k) + ...,
\end{equation}
where $C$ contains field-independent terms, $F_1(\tilde U_k)$ and 
$F_2(\tilde U_k^2)$ represent the linear and quadratic terms in the 
periodic part of the potential. Deriving the WH--RG equation 
(\ref{Uflow}) with respect to one of the field variable and
inserting the expanded form of the potential into it, 
the WH--RG equation (\ref{Uflow}) becomes 
\begin{equation}
\label{approx_wh}
(2 + k \, \partial_k) \frac{d}{d\varphi_1}
{\tilde U}_k(\varphi_1,...,\varphi_N) = 
- \frac{1}{4\pi} \, \, \,  
\frac{\frac{d}{d\varphi_1} [F_1(\tilde U_k) + \ord{\tilde U_k^2}]}
{C + F_1(\tilde U_k) + \ord{\tilde U_k^2}}.
\end{equation}
Since the constant term $C$ is field-independent, 
Eq.~(\ref{approx_wh}) can be rewritten as
\begin{equation}
\label{approx2_wh}
(2 + k \, \partial_k) \frac{d}{d\varphi_1}
{\tilde U}_k(\varphi_1,...,\varphi_N) = 
- \frac{1}{4\pi} \, \, \,  
\frac{\frac{d}{d\varphi_1} 
\left[ \frac{F_1(\tilde U_k) + \ord{\tilde U_k^2}}{C} \right]}
{1 + \frac{F_1(\tilde U_k) + \ord{\tilde U_k^2}}{C}}.
\end{equation}
The mass-corrected UV-RG equation can be achieved by linearizing 
Eq. (\ref{approx2_wh}) and reads as 
\begin{equation}
\label{uv_wh}
(2 + k \, \partial_k) {\tilde U}_k(\varphi_1,...,\varphi_N) 
\approx - \frac{1}{4\pi} \, \,  
\frac{F_1(\tilde U_k)}{C}.
\end{equation}

%--------------------------------------------------------------------
% N=2 LSG
%--------------------------------------------------------------------
\section{Flavour-doublet layered sine-Gordon model}
\label{N=2}

%--------------------------------------------------------------------
% N=2 Definition 
%--------------------------------------------------------------------
\subsection{Definition and rotation}
\label{defN=2}

In this Section we discuss the rotation of the LSG model with $N=2$
layers and apply the mass-corrected UV WH--RG method in order to map 
out the phase structure of the model. The ansatz for the blocked 
potential should preserve all symmetries of the original model at 
the UV cutoff scale $k=\Lambda$ and should be rich enough to contain
all the interactions which are generated during the RG flow. Therefore, 
the specialization of Eq.~(\ref{clhere}) to the case of two layers yields 
\begin{eqnarray}
\label{lsg_periodic}
& {\mathcal L}_{\rm 2LSG} =
{\frac12} \, \sum_{i=1}^2 (\partial \varphi_i)^2 + 
+ {\frac12} \, J(\varphi_1 -\varphi_2)^2 
\nonumber\\ 
&+ \sum_{n,m=0}^{\infty} 
\left[u_{nm} \cos(n\beta \, \varphi_1)\cos(m\beta \, \varphi_2) +
v_{nm} \sin(n\beta \, \varphi_1)\sin(m\beta \, \varphi_2)
\right]\,,
\end{eqnarray}
where the Fourier decomposition of the periodic part has a general 
form. All couplings $u_{nm}$ and $v_{nm}$ are dimensionful. For 
$N=2$-layers the mass eigenvalues are $0, 2J$. The particular choice 
of $\beta =2\sqrt{\pi}$ for the LSG represents the bosonized version 
of the two-flavour massive Schwinger model.

In order to emphasis the symmetries of the LSG model we now apply a 
rotation of the field variables described in Ref.~\cite{plb},
\begin{equation}
\label{rotation2}
\varphi_1 \to \frac{\alpha_1 + \alpha_2}{\sqrt{2}}\,,
\qquad
\varphi_2 \to \frac{\alpha_1 - \alpha_2}{\sqrt{2}}\,,
\end{equation}
where the periodic part of the blocked potential
\begin{eqnarray}
U_k(\varphi_1, \varphi_2) = \sum_{n,m=0}^{\infty} & 
\left[u_{nm} \cos(n\beta \, \varphi_1)\cos(m\beta \, \varphi_2) \right. 
\nonumber \\
& \left. + v_{nm} \sin(n\beta \, \varphi_1)\sin(m\beta \, \varphi_2)\right]
\end{eqnarray}
has the following rotated form 
\begin{eqnarray}
& U_k(\alpha_1, \alpha_2) =
\sum_{n,m=0}^{\infty} 
\frac{u_{nm} + v_{nm}}{2} \,
\cos\left[(n-m) \frac{\beta}{\sqrt{2}} \, \alpha_1\right]\,
\cos\left[(n+m) \frac{\beta}{\sqrt{2}} \, \alpha_2\right]\,
\nonumber\\
& \quad  + \sum_{n,m=0}^{\infty} \frac{u_{nm} - v_{nm}}{2} \,
\cos\left[(n+m) \frac{\beta}{\sqrt{2}} \, \alpha_1\right]\,
\cos\left[(n-m) \frac{\beta}{\sqrt{2}} \, \alpha_2\right]\,
\nonumber\\
& \quad  - \sum_{n,m=0}^{\infty} \frac{u_{nm} + v_{nm}}{2} \,
\sin\left[(n-m) \frac{\beta}{\sqrt{2}} \, \alpha_1\right]\,
\sin\left[(n+m) \frac{\beta}{\sqrt{2}} \, \alpha_2\right]\,
\nonumber\\
& \quad  + \sum_{n,m=0}^{\infty} \frac{v_{nm} - u_{nm}}{2} \,
\sin\left[(n+m) \frac{\beta}{\sqrt{2}} \, \alpha_1\right]\,
\sin\left[(n-m) \frac{\beta}{\sqrt{2}} \, \alpha_2\right]\,.
\end{eqnarray}
The general form of the rotated periodic 
potential reads as 
\begin{eqnarray}
\label{rotatedLN=2}
U_k =
\sum_{n,m=0}^{\infty} 
\left[f_{nm} \cos(n b\, \alpha_1)\cos(m b\, \alpha_2) 
+ h_{nm} \sin(n b\, \alpha_1)\sin(m b\, \alpha_2)\right]\,,
\end{eqnarray}
where the rotated frequency $b = \beta/\sqrt{2}$.
Some identifications read $f_{02} = \hf (u_{11} + v_{11})$, 
$f_{20} = \hf (u_{11} - v_{11})$ and $f_{11} = u_{01} + u_{10}$.
Finally, the rotated Lagrangian reads  
\begin{eqnarray}
& {\mathcal L}_{\rm 2LSG} =
{\frac12} \, (\partial \alpha_1)^2 + 
{\frac12} \, (\partial \alpha_2)^2
+ \frac12 \, M_2^2 \alpha_2^2 
+ U_k(\alpha_1, \alpha_2) \,.
\end{eqnarray}
Please notice that the field $\alpha_1$ has no explicit mass term
but for $\alpha_2$ the explicit mass $M_2^2= J/2$ breaks the periodicity.
Therefore, the model is disentangled into a ``periodic'' mode 
$\alpha_1$ and a non-periodic field $\alpha_2$.

%-------------------------------------------------------------------
% N=2 WH--RG  
%-------------------------------------------------------------------
\subsection{Wegner--Houghton RG approach to the rotated 
flavour-doublet model}
\label{WHN=2}

The specialization of the dimensionless WH--RG equation 
(\ref{dimless_wh}) for two layers can be written as 
\begin{equation}
\label{EQWHN=2}
(2 + k \, \partial_k) {\tilde V}_k(\alpha_1,\alpha_2) = 
- \frac{k^2}{4\pi}
\ln \left([1 + {\tilde V}^{11}_k] [1 + {\tilde V}^{22}_k] -
[\tilde V^{12}_k]^2 \right)\,,
\end{equation}
where ${\tilde V}^{ij}_k= \partial_{\alpha_i}
\partial_{\alpha_j} {\tilde V}_k (\alpha_1,\alpha_2)$.
Starting with a general form for the dimensionless rotated blocked
potential for the flavour-doublet LSG model 
\begin{eqnarray}
{\tilde V}_k= \frac12 \, {\tilde M}_2^2 \alpha_2^2 +
\sum_{n,m=0}^{\infty}&
\left[{\tilde f}_{nm} \cos(n b\, \alpha_1)\cos(m b\, \alpha_2) \right.
\nonumber \\
&\left. + {\tilde h}_{nm} \sin(n b\, \alpha_1)\sin(m b\, \alpha_2)\right]\,,
\end{eqnarray}
where $\tilde f_{nm} = k^{-2} f_{nm}$ and $\tilde h_{nm} = k^{-2} h_{nm}$
are the dimensionless coupling constants and using the ``mass-corrected'' 
UV approximation of the WH--RG equation (\ref{EQWHN=2}) which based 
on Eq.~(\ref{uv_wh}), this reduces to a set of uncoupled differential 
equations for the coupling parameters of the model 
%
%\begin{subequations}
\begin{eqnarray}
& (2+ k\partial_k) {\tilde f}_{nm}(k) = 
\frac{1}{4\pi} \, 
\frac{k^2 m^2 b^2 + (k^2 + M_2^2) n^2 b^2}{(k^2 + M_2^2)} 
{\tilde f}_{nm} \,,
\nonumber \\ 
& (2+ k\partial_k) {\tilde h}_{nm}(k) =
\frac{1}{4\pi} \, 
\frac{k^2 m^2 b^2 + (k^2 + M_2^2) n^2 b^2}{(k^2 + M_2^2)}
{\tilde h}_{nm}\,,
\end{eqnarray}
%\end{subequations}
% 
where the dimensionful mass $M^2_2$ is scale-independent. 
The UV approximated RG flow equations are decoupled, 
and their solution can be obtained analytically:
%
%\begin{subequations}
\begin{eqnarray}
\label{solut}
& {\tilde f}_{nm}(k) = {\tilde f}_{nm} (\Lambda)
\left(\frac{k^2 + M_2^2}{\Lambda^2 + M_2^2}\right)^{\frac{ m^2 b^2}{8\pi}}
\left(\frac{k}{\Lambda}\right)^{-2+\frac{ b^2 n^2}{4\pi}} \,,
\nonumber \\ 
& {\tilde h}_{nm}(k) = {\tilde f}_{nm} (\Lambda)
\left(\frac{k^2 + M_2^2}{\Lambda^2 + M_2^2}\right)^{\frac{ m^2 b^2}{8\pi}}
\left(\frac{k}{\Lambda}\right)^{-2+\frac{ b^2 n^2}{4\pi}} \,.
\end{eqnarray}
%\end{subequations}
% 
Here ${\tilde f}_{nm}(\Lambda)$ and ${\tilde h}_{nm}(\Lambda)$
are the initial conditions at the UV cutoff $k = \Lambda$. Using the 
solution (\ref{solut}), one can read off the IR scaling of the various
Fourier amplitudes. In the IR limit ($k\to 0$) all the purely non-periodic
modes ($n=0$) become relevant, i.e.~$\tilde f_{0m} \propto k^{-2}$.
The periodic modes $\tilde f_{nm}$ and $\tilde h_{nm}$
(with $n>0$) may be relevant or irrelevant,
depending on the value of $b^2$. If $b^2 > 8\pi$, the RG flow of all 
the periodic modes tends to zero, and if $b^2 < 8\pi$, there is 
at least one mode which becomes relevant in the IR limit. 
We recall that $b^2 = \beta^2/2$. Therefore, the critical value which
separates the two scaling regime of the original LSG model is 
$\beta_c^2=16\pi$. We would like to remind that the LSG model with $N=1$ 
layer is the 2D-SG model with $\beta_c^2 = 8\pi$. So, in case of $N=2$ 
layers the critical value is increased compared to that of the 2D-SG model.

%------------------------------------------------------------
% N=3 LSG
%-------------------------------------------------------------
\section{Flavour-triplet layered sine-Gordon model}
\label{N=3}

%-----------------------------------------------------------
% N=3 Definition 
%-----------------------------------------------------------
\subsection{Definition and rotation}
\label{defN=3}

The Lagrangian of the LSG model with $N=3$ layers can be written as
\begin{eqnarray}
\label{3lsg}
{\mathcal L}  =&
{1\over 2} \, \sum_{i=1}^3 (\partial \varphi_i)^2 + 
{1\over 2} \, J \sum_{i=1}^2 (\varphi_{i+1} -\varphi_i)^2 
\nonumber\\ 
& + \sum_{n,m,l=-\infty}^{\infty} 
w_{nml} \,
\exp\left({\rm i}\,n \beta\, \varphi_1\right)\,
\exp\left({\rm i}\,m \beta\, \varphi_2\right)\,
\exp\left({\rm i}\,l \beta\, \varphi_3\right)\,.
\end{eqnarray}
The parameters $w_{nml}$ of the Fourier decomposition are 
dimensionful quantities. The mass matrix with eigenvalues 
($0, J, 3J$) reads explicitly
\begin{equation}
\label{prop_3lsg}
{\underline{\underline M}}^2= 
\left( \begin{array}{ccc} 
J  &\hspace*{0.5cm} -J &\hspace*{0.5cm}  0\\
-J &\hspace*{0.5cm} 2J &\hspace*{0.5cm} -J\\
0  &\hspace*{0.5cm} -J &\hspace*{0.5cm}  J\\
\end{array} \right) \,.
\end{equation}
The explicit form of the rotation of the field variables in 
the case of three layers has the following structure, 
%
%\begin{subequations}
\begin{eqnarray}
\label{trafo3layer}
\varphi_1 \to& \frac{\alpha_1}{\sqrt{3}} - \frac{\alpha_2}{\sqrt{2}}  
+ \frac{\alpha_3}{\sqrt{6}} \,,
\nonumber \\[1ex]
\varphi_2 \to& \frac{\alpha_1}{\sqrt{3}}   
- \frac{\sqrt{2}\alpha_3}{\sqrt{3}} \,,
\nonumber \\[1ex]
\varphi_3 \to& \frac{\alpha_1}{\sqrt{3}} + \frac{\alpha_2}{\sqrt{2}}  
+ \frac{\alpha_3}{\sqrt{6}} \,.
\end{eqnarray}
%\end{subequations}
%
For illustrative purposes the transformation of the periodic 
part of the bare potential is discussed by taking into account 
only the fundamental modes. In this case the bare potential has 
a flavour symmetry ($\varphi_1 \longleftrightarrow \varphi_3$) and 
reads as
\begin{equation}
\label{fundansatz}
U(\varphi_1, \varphi_2, \varphi_3) =
u  \cos(\beta \, \varphi_1) +
u_2 \cos(\beta \, \varphi_2) +
u  \cos(\beta \, \varphi_3) \,,
\end{equation}
where $w_{100} = w_{001} \equiv u/2$ and $w_{010}\equiv u_{2}/2$ 
is introduced. Applying the rotation~(\ref{trafo3layer}) on the 
periodic potential (\ref{fundansatz}) the transformed potential is
\begin{eqnarray}
& U(\alpha_1, \alpha_2, \alpha_3) =
u_2 \,
\cos\left(\frac{\beta}{\sqrt{3}} \, \alpha_1\right)\,
\cos\left(\frac{2\beta}{\sqrt{6}} \, \alpha_3\right)\,
\nonumber\\
& \quad  + 2 u \,
\cos\left(\frac{\beta}{\sqrt{3}} \, \alpha_1\right)\,
\cos\left(\frac{\beta}{\sqrt{2}} \, \alpha_2\right)\,
\cos\left(\frac{\beta}{\sqrt{6}} \, \alpha_3\right)\,
\nonumber\\
& \quad  + 2 u \,
\sin\left(\frac{\beta}{\sqrt{3}} \, \alpha_1\right)\,
\cos\left(\frac{\beta}{\sqrt{2}} \, \alpha_2\right)\,
\sin\left(\frac{\beta}{\sqrt{6}} \, \alpha_3\right)\,
\nonumber\\
& \quad  + 2 u_2 \,
\cos\left(\frac{\beta}{\sqrt{3}} \, \alpha_1\right)\,
\sin\left(\frac{\beta}{\sqrt{2}} \, \alpha_2\right)\,
\sin\left(\frac{\beta}{\sqrt{6}} \, \alpha_3\right)\,.
\end{eqnarray}
In general, the rotated form of the blocked periodic potential reads as
\begin{equation}
\label{rotatedLN=3}
U_k =
\sum_{n,m,l=-\infty}^{\infty} 
j_{nml} \,
\exp\left(\frac{{\rm i}\,n \beta}{\sqrt{3}}\, \alpha_1\right)\,
\exp\left(\frac{{\rm i}\,m \beta}{\sqrt{2}}\, \alpha_2\right)\,
\exp\left(\frac{{\rm i}\,l \beta}{\sqrt{6}}\, \alpha_3\right)\,,
\end{equation}
where the $j_{nml}$ are the transformed expansion coefficients. The 
new frequency for the periodic mode $\alpha_1$ is $b_1 = \beta/\sqrt{3}$. 
For the two non-periodic modes (with explicit mass terms), the 
transformed frequencies read $b_2 = \beta/\sqrt{2}$ and 
$b_3 = \beta/\sqrt{6}$. After rotation the Lagrangian of the $N=3$
layer model is
\begin{eqnarray}
& {\mathcal L}_{\rm 3LSG} =
\sum_{i=1}^3 {1\over 2} \, (\partial \alpha_i)^2 
+ \frac12 \, M_2^2 \alpha_2^2  
+ \frac12 \, M_3^2 \alpha_3^2 
+ U_k(\alpha_1, \alpha_2, \alpha_3) \,,
\end{eqnarray}
with mass eigenvalues $M_2^2 = J$ and $M_3^2 = 3J$. Like in the 2-layer 
case we have decomposed the three-layer model into one periodic 
mode $\alpha_1$ and two non-periodic fields $\alpha_2$,  $\alpha_3$.

%-------------------------------------------------------------------
% N=3 WH--RG
%-------------------------------------------------------------------
\subsection{Wegner--Houghton RG approach to the rotated 
flavour-triplet model}
\label{WHN=3}

\begin{figure}[ht]
\begin{center}
\epsfig{file=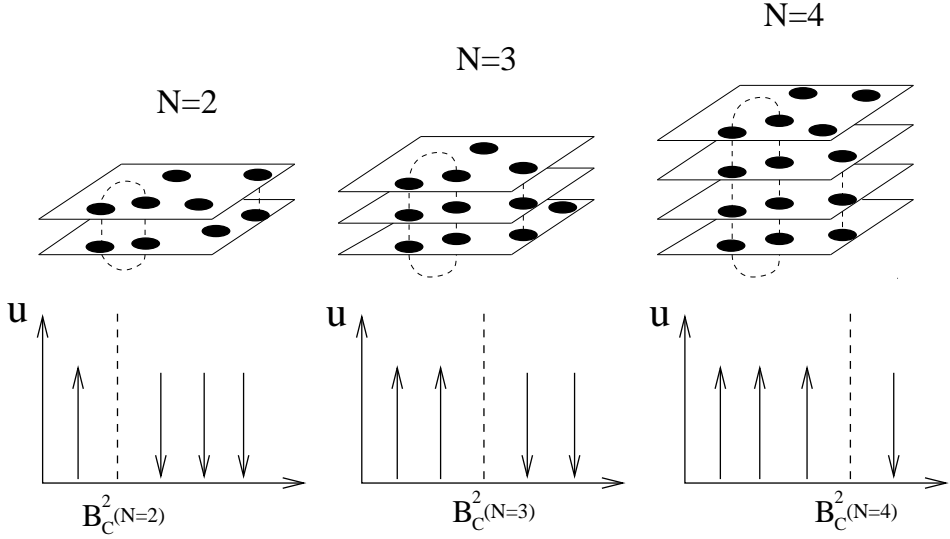,width=0.8\linewidth}
\caption{\label{fig1} We illustrate the schematic RG trajectories 
of the multi-layer sine-Gordon model with $N=2,3,4$ layers in the plane 
$(B^2\equiv\beta^2,u\equiv \tilde u_{01})$ and the shift of the critical 
value $B_c^2(N) \equiv \beta^2_c (N) = 8 N\pi$. Each layer corresponds 
to a sine-Gordon model which are coupled by the coupling $J$. The solid 
discs represent the topological excitation of the layered system.
}
\end{center}
\end{figure}

We generalize the treatment discussed in Sec.~\ref{WHN=2} to the
case of three layers by repeating the same steps as in Sec.~\ref{WHN=2}. 
The rotated dimensionless blocked potential for the flavour-triplet 
LSG model is
\begin{eqnarray}
{\tilde V}_k(\alpha_1, \alpha_2, \alpha_3) 
= \frac12 \, {\tilde M}_2^2 \alpha_2^2 
+ \frac12 \, {\tilde M}_3^2 \alpha_3^2 
\nonumber \\
+ \sum_{n,m,l=-\infty}^{\infty} {\tilde j}_{nml} \,
\exp\left({\rm i}\,n b_1\, \alpha_1\right)\,
\exp\left({\rm i}\,m b_2\, \alpha_2\right)\,
\exp\left({\rm i}\,l b_3\, \alpha_3\right).
\end{eqnarray}
The dimensionless WH--RG equation in $d=2$ dimensions,
for three fields $\alpha_{1,2,3}$, reads
\begin{eqnarray}
\label{EQWHN=3}
(2 + k\,\partial_k) {\tilde V}_k = - \frac{k^2}{4\pi} \, 
\ln &\left(
[1 + \tilde V^{11}_k] [1 + \tilde V^{22}_k] [1 + \tilde V^{33}_k]  \right.
- [1 + \tilde V^{22}_k] [\tilde V^{13}_k]^2 
\nonumber\\
& - [1+ \tilde V^{33}_k] [\tilde V^{12}_k]^2 
- [1 + \tilde V^{11}_k] [\tilde V^{23}_k]^2
\nonumber\\
& \left. + [\tilde V^{12}_k] [\tilde V^{23}_k] [\tilde V^{31}_k] 
+ [\tilde V^{13}_k] [\tilde V^{21}_k] [\tilde V^{32}_k]
\right) \,.
\end{eqnarray}
where $\tilde V^{ij}_k= \partial_{\alpha_i}
\partial_{\alpha_j}\tilde V_k (\alpha_1,\alpha_2,\alpha_3) $ is
the second derivative of the dimensionless blocked potential
with respect to the field variables. In order to obtain reliable
UV scaling laws which can be used to determine the phase diagram
of the LSG model with $N=3$ layers one has to use the mass-corrected 
UV approximation for the WH--RG equation (\ref{EQWHN=3}). This reduces 
to a set of uncoupled differential equations for the coupling 
constants of the model 
\begin{eqnarray}
& (2+ k\partial_k) {\tilde j}_{nml}(k) = \frac{1}{4\pi} 
\left( n^2 b_1^2 + \frac{k^2 m^2 b_2^2}{k^2 + M_2^2} + 
\frac{k^2 l^2 b_3^2}{k^2 + M_3^2} \right)
{\tilde j}_{nml} \,.
\end{eqnarray}
The solution can be obtained analytically,
\begin{eqnarray}
& {\tilde j}_{nml}(k) = {\tilde j}_{nml} (\Lambda)
\left(\frac{k}{\Lambda}\right)^{-2+\frac{ b_1^2 n^2}{4\pi}}
\left(\frac{k^2 + M_2^2}{\Lambda^2 + M_2^2}\right)^{\frac{m^2 b_2^2}{8\pi}}
\left(\frac{k^2 + M_3^2}{\Lambda^2 + M_3^2}\right)^{\frac{l^2 b_3^2}{8\pi}} \,,
\end{eqnarray}            
where ${\tilde j}_{nml}(\Lambda)$ is the initial condition at the 
UV cutoff $k = \Lambda$. In the IR regime ($k\to 0$), the pure 
non-periodic modes are relevant (increasing) coupling constants 
$\tilde j_{0ml} \propto k^{-2}$ independently of $b_i^2$.
The periodic modes $\tilde j_{nml}\,, n>0$ are found to be relevant 
or irrelevant couplings depending on the value of $b_1^2$. 
If $b_1^2 > 8\pi$, the RG flow of all the periodic modes tend to zero, 
and if $b_1^2 < 8\pi$, they become relevant in the IR limit. 
We recall that $b_1^2 = \beta^2/3$. Therefore, the critical value for
the original 3-layer LSG model is $\beta^2_c = 24\pi$. This suggest
that the critical value of the parameter which separates the 
two phases of LSG model with $N$-layers depends on the number of layers,
$\beta_c^2 = N 8\pi$ (see Fig.~\ref{fig1}). 
This is in agreement with the perturbative results
obtained in Ref.~\cite{plb}.

%
% Fig. 1
%

%-------------------------------------------------------------------
% 3D-SG model 
%-------------------------------------------------------------------
\section{3D-SG model}
\label{3dsg}

\begin{figure}[ht]
\begin{center}
\epsfig{file=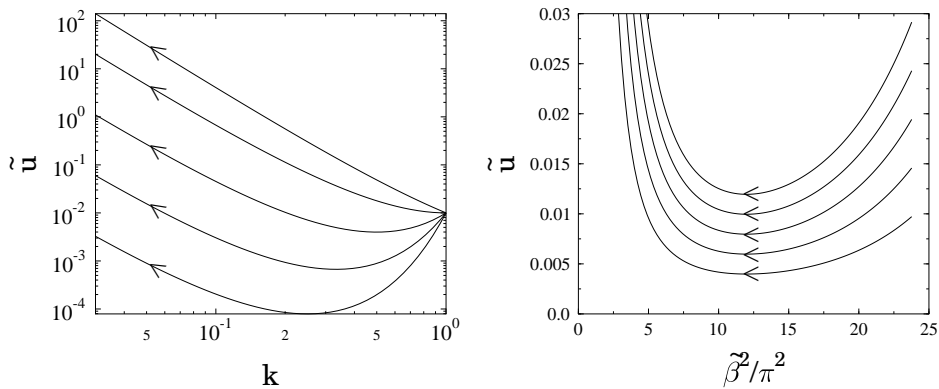,width=0.8\linewidth}
\caption{\label{fig2} On the left panel, the scaling of the 
dimensionless coupling constant $\tilde u(k)$ of the 3D--SG 
model is represented graphically for various initial values 
for the dimensionless parameter 
$\tilde \beta^2(\Lambda=1) = 4\pi^2, 12\pi^2, 24\pi^2, 36\pi^2, 48\pi^2$
(from top to bottom).
In the IR limit ($k\to 0$) the coupling constant $\tilde u$ always 
becomes a relevant (increasing) parameter independently of 
$\tilde \beta^2(\Lambda)$ which demonstrates that the 3D--SG model has 
a single phase in the local potential approximation. On the
right panel, the RG flow diagram of the 3D--SG model also illustrates the
existence of the single phase of the model and the infinite value of
$\beta^2_{\rm c}$. The arrows indicate the direction of the flow.
}
\end{center}
\end{figure}

Since in the continuum limit $N\to\infty$ the $N$-layer LSG model can
be considered as the discretized version of the 3D-SG model, one can
clarify the previously obtained layer-dependence of the critical 
parameter $\beta_c^2 = N 8 \pi$ by investigating the phase structure
of the 3D-SG model. One might expect, that for $N\to\infty$ the phase 
structure of the LSG model recovers that of the 3D-SG model.
The action for the 3D-SG model reads as
\begin{equation}
\label{3dsg_def}
S = \int d^3 r \left[ {\frac12} (\partial_{\mu} \varphi)^2 
+ u_{3D} \cos(\beta_{3D}\varphi)\right],
\end{equation}
where $\beta_{3D}$, $u_{3D}$ are the dimensionful parameters of the 
theory. The corresponding dimensionless quantities are 
$\tilde\beta^2 = k \beta^2_{3D} $ and $\tilde u = k^{-3} u_{3D}$. 
The WH--RG approach to the 3D-SG model has been developed and discussed 
in Ref.~\cite{sg3}. Using the local potential approximation, the 
dimensionless WH--RG equation reads as
\begin{equation}
\label{3dsg_wh}
(3 - \frac{1}{2} \tilde\varphi \partial_{\tilde\varphi} 
+ k \partial_k) \tilde V_k(\tilde\varphi) =
- \frac{1}{4 \pi^2} 
\ln\left( 1 +  \partial^2_{\tilde\varphi}\tilde V_k(\tilde\varphi)\right).
\end{equation}
The UV approximation for Eq.~(\ref{3dsg_wh}) can be achieved by
the linearization of the logarithm around the Gaussian fixed point
and results in 
\begin{eqnarray}
(3 + k\partial_k) \tilde u(k) = \frac{1}{4 \pi^2}\,
\tilde\beta^2(k) \, \tilde u(k),
\nonumber \\
k\partial_k \tilde\beta^2(k) = \tilde\beta^2(k).
\end{eqnarray}
with the solution
\begin{eqnarray}
\label{3dsg_sol}
{\tilde u}(k) = {\tilde u}(\Lambda)  
\left(\frac{k}{\Lambda}\right)^{-3} 
\exp \biggl\{ \frac{\tilde \beta^2(\Lambda)}{4\pi^2} 
\biggl[ \biggl(\frac{k}{\Lambda}\biggr) - 1 \biggr] \biggr\}
\nonumber \\
{\tilde \beta^2}(k) = {\tilde \beta^2}(\Lambda)  
\left(\frac{k}{\Lambda}\right) 
\end{eqnarray}
where $\tilde \beta(\Lambda)$ and ${\tilde u}(\Lambda)$ are the bare
values of the couplings. In the IR limit, if $k\to 0$ the coupling
constant $\tilde u(k)$ always becomes a relevant parameter 
($\tilde u\to \infty$) independently of $\tilde \beta^2$, 
(see Fig.~\ref{fig2}).
Therefore, the 3D-SG model has only a single phase within the LPA.

%
% Fig. 2
%

On the one hand, in case of the LSG model for the bulk limit 
($N \to \infty$), the critical value which separates the two phases of 
the layered model becomes infinitely large ($\beta^2_{\rm c} \to \infty$) 
and the model has only a single phase. On the other hand, in this 
continuum limit, the multi-layer model can be considered as the 
discretized version of the 3D-SG model which has been shown to have a 
single phase, within in the local potential approximation 
(see Fig.~\ref{fig2} and Ref.~\cite{sg3}). We conclude that the 
latter observation is entirely consistent with the infinite value 
of $\beta^2_{\rm c}$ in the continuum limit.

%------------------------------------------------------------------
\section{Summary}
%------------------------------------------------------------------
\label{sum} 

The phase structure of the layered sine-Gordon (LSG) model with 
$N$-layers has been analyzed in terms of a non-perturbative 
renormalization group (RG) treatment with a sharp momentum cutoff. 
The LSG model consists of $N$ coupled SG models each of which 
corresponds to a specific layer. The coupling between the layers 
is described by a quadratic term which can be considered as a mass 
term. All the Lagrangians studied in the paper have the general 
structure~(\ref{clb}). The case with a mass matrix that has exactly 
one non-vanishing mass eigenvalue has been discussed in detail.

The LSG model has relevance both in high-energy and, perhaps 
even more importantly, in low-temperature
physics. The $N$-layer SG model is the bosonised version of the 
$N$-flavour Schwinger model, and the double-layer SG model has been 
used to describe the vortex properties of high transition 
temperature superconductors~\cite{pierson_lsg,philmag}.

Previously, models of this type, with up to two layers, were 
analyzed in terms of the Wegner--Houghton renormalization group (WH--RG) 
method~\cite{NaEtAl2005,sg3}. We here describe the 
generalization of the RG analysis for
$N$ layers, which has allowed us to consider the dependence of the phase
structure on the number of the layers. In order to be able to compare 
the results of our RG analysis to that of the perturbative treatment 
performed for the LSG model~\cite{plb}, we have performed a rotation 
of the fields before applying the WH--RG method.

It has been demonstrated that the LSG model undergoes a Kosterlitz--Thouless 
type phase transition where the critical value which separates the
two phases of the model depends on the number of layers, 
$\beta_c^2= N 8 \pi$. Therefore, the transition ``temperature'' 
in the layered case was found to differ from a one-layer ``pure'' 
sine-Gordon model. Furthermore, we have shown 
(see also~\cite{plb}) that this conjecture finds a 
natural explanation after a suitable linear transformation of 
the field variables, which corresponds to a rotation in the 
internal space towards a frame in which the mass matrix is diagonal.

The LSG model in the continuum limit $N\to\infty$ has been shown 
to be considered as the discretized version of the three-dimensional 
SG model which has a single phase within the local potential approximation. 
The infinite critical value of the parameter $\beta_c^2$ for the LSG 
model in the continuum limit is entirely 
consistent with the latter observation.

%------------------------------------------------------------------
% Acknowledgments
%------------------------------------------------------------------
\section*{Acknowledgments}

The author acknowledges the numerous fruitful discussions with 
U. D. Jentschura, J. Zinn-Justin, S. Nagy, K. Sailer, K. Vad and 
S. M{\'e}sz{\'a}ros and the warm hospitality during a visit to 
the Max--Planck--Institute for Nuclear Physics (Heidelberg).

%------------------------------------------------------------------
% References
%------------------------------------------------------------------

\end{document}